\documentclass[twoside]{photon2007} 
\usepackage[latin1]{inputenc} 
\usepackage[dvips]{graphicx,epsfig,color} 
\usepackage{wrapfig,rotating} 
\usepackage{amssymb,amsmath,array} 
\usepackage{multirow}  
\input{symblibrary.tex} 
\begin{document} 
%%%%   Paper title goes here  %%%%%%%%%%%%%% 
  \title{Latest Results on Bottom Spectroscopy and Production with CDF} 
%%  
%*********************************************************************** 
% AUTHORS INFORMATION AREA 
%*********************************************************************** 
  \author{ 
    Igor V. Gorelov  
% Optional short acknowledgment: remove next line if non-needed 
    \thanks{This talk~\cite{url} has been presented on behalf of 
            the CDF Collaboration at a conference ``Photon 2007''.} 
% DO NOT MODIFY THE FOLLOWING '\vspace' ARGUMENT 
    \vspace{.3cm}\\ 
% Addresses and institutions (remove "1- " in case of a single institution) 
    Department of Physics and Astronomy, \\ 
    MSC07 4220, University of New Mexico, \\  
    800 Yale Blvd. NE, \\ 
    Albuquerque, NM 87131, USA 
} 
%%*********************************************************************** 
% END OF AUTHORS INFORMATION AREA 
%*********************************************************************** 
% 
\maketitle 
\begin{abstract} 
  Using data collected with the CDF Run II detector, new 
  measurements on bottom production cross-sections are presented. The 
  latest achievements in bottom hadron spectroscopy are discussed. 
  The results are based on a large sample of semileptonic and hadronic 
  decays of bottom states made available by triggers based on the 
  precise CDF tracking system. 
\end{abstract} 
\section{First Observation of the Baryons \( \mathbf{\Sigb} \) and \( \mathbf{\Sigbst} \) in CDF} 
\begin{figure}[hbt] 
  %\centerline{ 
  \includegraphics[width=1.0\columnwidth]{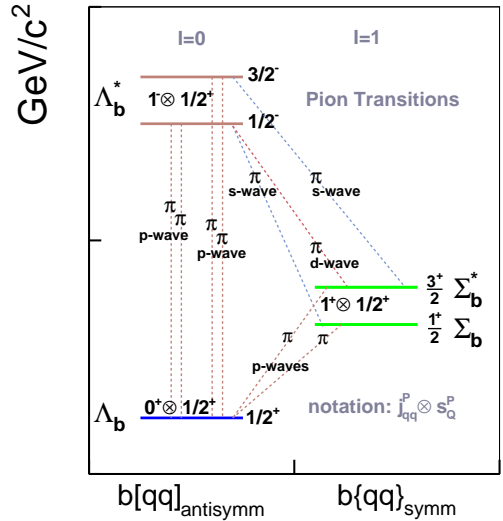} 
  %} 
  \caption{The low lying $\Sigma$- and $\Lambda$- like \b-baryons and their strong  
           decays with pion emissions.} 
  \label{fig:pitrans}  
  \end{figure}

\begin{figure}[h] 
  \hspace{-0.15in}\includegraphics[width=1.0\columnwidth]{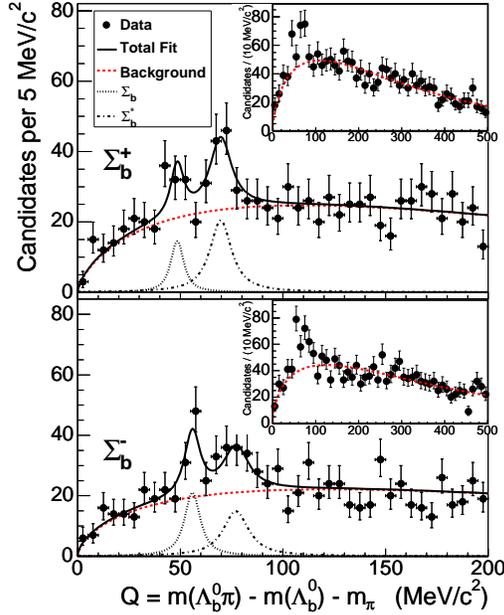} 
  \caption{The experimental mass difference 
           spectra~\cite{exp:sigb-prl} for the candidates of both 
           charged partners, \Sgbstpm. Double peak signatures are 
           observed in every case.} 
  \label{fig:sigb}  
  \end{figure}

  The bottom \Sgbst states decay strongly into \Lb by emitting soft pion  
  as shown in Figure~\ref{fig:pitrans}. 
% 
%\begin{wrapfigure}{r}{1.0\columnwidth} 
%\end{wrapfigure} 
% 
  Our results are based on data collected with the \cdf2 
  detector~\cite{cdf:det} and corresponding to an integrated 
  luminosity of $\sim1.1\invfb$.  The trigger used in this study is 
  based on displaced tracks.  It reconstructs with the central tracker 
  a pair of \( \pt\gsim2.0\gevc \) tracks at Level 1 and enables secondary vertex 
  selection at Level 2 requiring each of these tracks to have impact 
  parameter measured by the CDF silicon detector SVX II larger than 
  120\mum. 
  The signals of \Sgbstpm states were sought in the 
  decay chain \( { \Sgbstpm\to\LambB\pi_{soft}^{\pm} } \),  
  \( {\LambB\to\Lcpim } \), \( { \Lc\to\pKpi } \)\footnote{Unless 
  otherwise stated all references to the specific charge combination 
  imply the charge conjugate combination as well.}. 
  To remove the contribution due to a mass resolution of each \Lb 
  candidate and to avoid absolute mass scale systematic uncertainties, 
  the \Sgbstpm candidates were reconstructed in the mass difference 
  {\it Q-value} spectra defined as  
  \( Q =  M(\LambB\pi_{soft}^{\pm})-M(\Lb)-\,M_{\rm PDG}(\pipm) \) 
  for every charge state of \Sgbstpm candidates. Here we assume also  
  that the width of the weakly decaying \Lb candidate is determined  
  by the corresponding detector mass resolution. 
  The fitted experimental spectra are shown at Figure~\ref{fig:sigb}, and   
  fit results are summarized in Tables~\ref{tab:sigb-mass}  
  and~\ref{tab:sigb-num}~~\cite{exp:sigb-prl}. 
  \begin{table}[h] 
  \centerline{ 
  \begin{tabular}{l l l}  
  \hline\hline 
  State & \multicolumn{1}{c}{ $Q$ or $\Delta_{\Sigbst}$ ($\mevcc$) } & \multicolumn{1}{c}{Mass ($\mevcc$) } \\ 
  \hline 
  $\Sigbp$ & $Q_{\Sigbp} = 48.5^{+2.0+0.2}_{-2.2-0.3}$ 	& $5807.8^{+2.0}_{-2.2}\pm 1.7$	\\ 
  $\Sigbm$ & $Q_{\Sigbm} = 55.9\pm 1.0\pm 0.2$		& $5815.2\pm 1.0\pm 1.7$ \\ 
  $\Sigbstp$ & \multirow{2}{*}{$\Delta_{\Sigbst} = 21.2^{+2.0+0.4}_{-1.9-0.3}$} & $5829.0^{+1.6+1.7}_{-1.8-1.8}$ \\ 
  $\Sigbstm$ & 	& $5836.4\pm 2.0^{+1.8}_{-1.7}$	\\ 
  \hline\hline 
  \end{tabular} 
  } 
  \caption{The masses resulting from the simultaneous fit 
           of both spectra~\cite{exp:sigb-prl}.} 
  \label{tab:sigb-mass} 
  \end{table} 
  \begin{table}[h] 
  \centerline{ 
  \begin{tabular}{c c c c}  
  \hline\hline 
  \multicolumn{4}{c}{Yields of the signals} \\ 
  \hline 
  $\Sigbp$	& $\Sigbm$ & $\Sigbstp$ & $\Sigbstm$  \\ 
  \hline 
  $32^{+13+5}_{-12-3}$ & $59^{+15+9}_{-14-4}$ & $77^{+17+10}_{-16-6}$ & $69^{+18+16}_{-17-5}$ \\ 
  \hline\hline 
  \end{tabular} 
  } 
  \caption{The fitted yields~\cite{exp:sigb-prl} of the identified 
           \Sgbstpm states. The combined significance of all four peaks 
           relative to the null hypothesis well exceeds 5 Gaussian 
           standard deviations.} 
  \label{tab:sigb-num} 
  \end{table} 
\section{ Observation and Mass Measurement of the Baryon \( \mathbf{\Xib} \) } 
\begin{figure}[hbt] 
  \includegraphics[width=1.0\columnwidth]{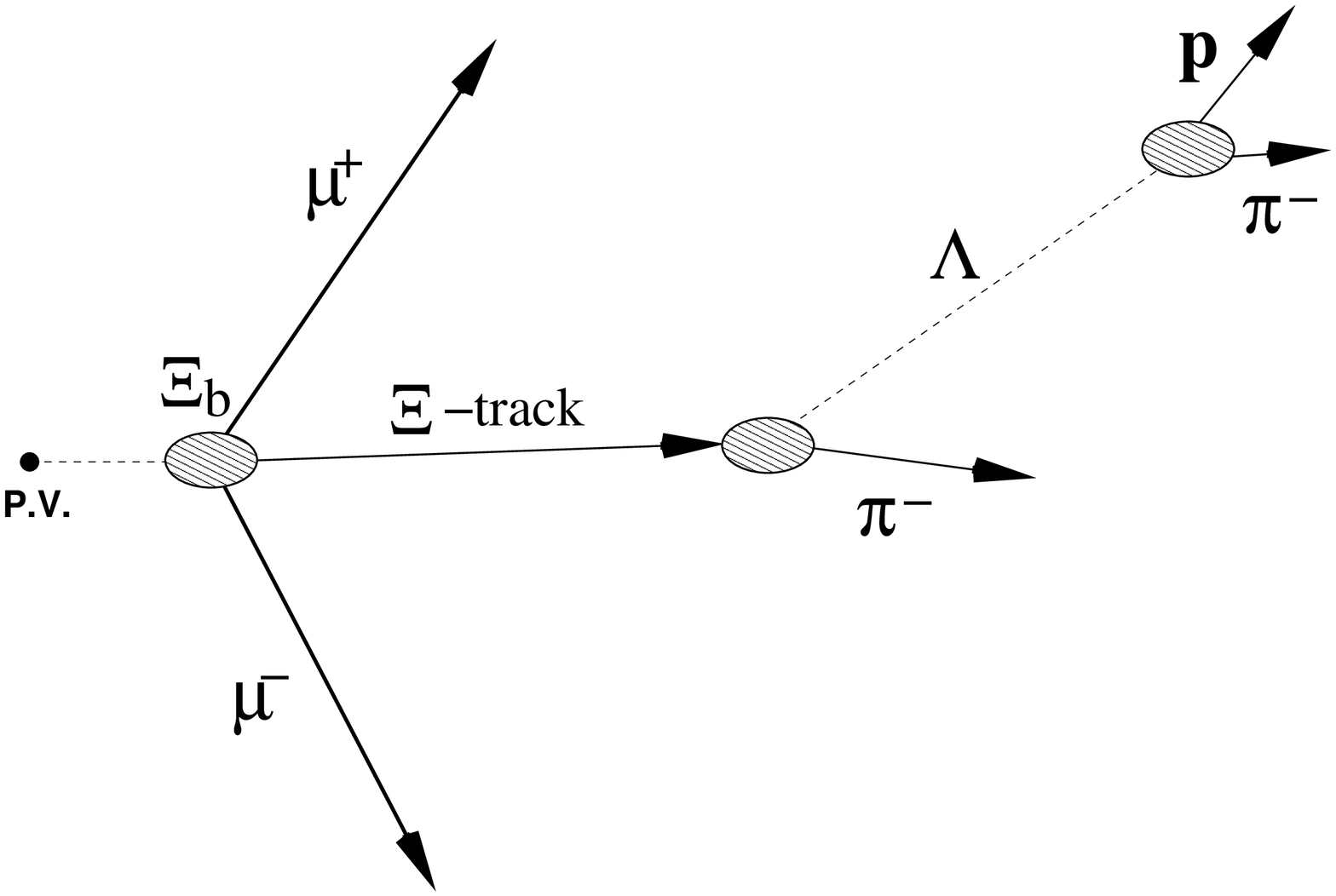} 
  \caption{Topology of the \( \Xibm\to\jpsi\Xism \) decay. } 
  \label{fig:xib-top}  
  \end{figure}

\begin{figure}[hbt] 
  \includegraphics[width=1.0\columnwidth]{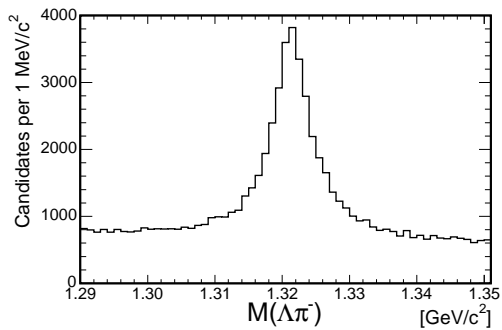} 
  \caption{The \( \Xism \) signal~\cite{exp:xib-prl} when the cascade 
           track has at least 2 hits in the CDF SVX II 
           tracker.} 
  \label{fig:xib-xis}  
  \end{figure}

\begin{figure}[hbt] 
  \includegraphics[width=1.0\columnwidth]{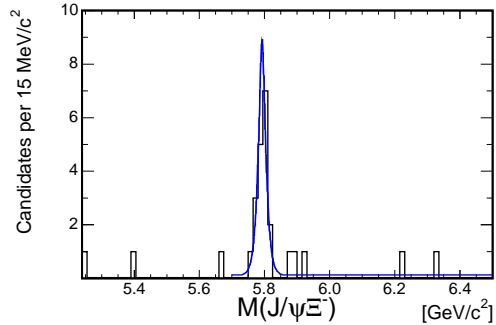} 
  \caption{The invariant mass distribution of \( \jpsi\Xism \) 
           candidates after optimized selection criteria have been applied. The 
           profile of the unbinned fit is superimposed. A clear 
           signal is observed~\cite{exp:xib-prl}. } 
  \label{fig:xib-signal}  
  \end{figure}

  The bottom cascade baryons \Xib consist of a single bottom quark, 
  one strange quark and one light quark.  Theoretical 
  predictions for these heavy baryons are outlined 
  in Table~\ref{tab:xib-th}~\cite{th:koerner}. 
  \begin{table}[b] 
  \centerline{ 
  \begin{tabular}{lllllll} 
  \hline\hline 
  State         & $b~sq$    & $J^{P}$    &$I_{3}$ & $j_{sq}$ & {\small M,\gevcc} \\  
  \hline 
  $\Xi_b^0$     & $b[su]$   & $1/2^{+}$  & 1/2    & 0        & $5.80$ \\ 
  $\Xi_b^-$     & $b[sd]$   & $1/2^{+}$  &-1/2    & 0        & $5.80$ \\  
  $\Xi_b^{0'}$  & $b\{su\}$ & $1/2^{+}$  & 1/2    & 1        & $5.94$ \\ 
  $\Xi_b^{-'}$  & $b\{sd\}$ & $1/2^{+}$  &-1/2    & 1        & $5.94$ \\  
  \hline\hline 
  \end{tabular} 
  } 
  \caption{Theoretical expectations for properties of bottom 
           cascade baryons containing a single \b- quark~\cite{th:koerner}.   
           The lowest 
           lying states have a light quark pair with momentum 
           $j_{sq}=0$ while the next ones have light quarks aligned 
           with $j_{sq}=1$. } 
  \label{tab:xib-th} 
  \end{table} 
  We consider the lowest lying \Xib states that decay weakly and the 
  $\Xib^{'}$ states that decay radiatively or strongly via 
  pion emission.  The \Xib candidates are reconstructed in the decay chain \( \Xibm\to\jpsi\Xism \)  
  with secondary states \( \jpsi\to\mumu \) and 
  \( \Xism\to\Lamb\pim\),  
  \( \Lamb\to\proton\pim \) (see Figure~\ref{fig:xib-top}) . 
  Since experiments with bubble chambers the strange cascade, given its long 
  decay path of $c\cdot\tau=4.91\cm$~\cite{pdg}, is identified as a 
  charged track with a 1-track decay vertex at the end formed by a 
  kinked soft pion track as shown at Figure~\ref{fig:xib-top}. 
  The subsequent $V^{0}$ decay vertex of the $\Lz$ is associated with the 
  1-track vertex and included in a two-vertex kinematic fit.  The 
  key technique in this analysis is the tracking algorithm 
  developed to reconstruct \( \Xism \) tracks leaving hits in the CDF silicon 
  vertex tracker SVX II. A finest tracking resolution~\cite{cdf:det} coupled with the 
  custom software provide a clean signal for \( \Xism \), see 
  Figure~\ref{fig:xib-xis}. 
  The analysis~\cite{exp:xib-prl} uses a data sample of  
  integrated \( \Lumi=1.9\invfb \) collected by the CDF dimuon trigger~\cite{cdf:det} 
  which saves events with two oppositely charged tracks reconstructed in the  
  CDF central tracker, matched to hits in the CDF muon chambers and 
  selected in the mass window \( M(\mumu)\in[2.7,4.0]\gevcc \) around 
  the mass of the \jpsi~\cite{pdg}. 
  The sample yields \( \sim15\times10^{6} \) \jpsi and \( \sim23500 \)  
  \Xism candidates. The final selection criteria for \Xibm candidates  
  have been studied using \( \sim31000 \) \B-mesons in the mode  
  \Bp\to\jpsi\Kp as a control sample assuming very similar decay kinematics. 
  The invariant mass of selected \( \jpsi\Xism \) candidates is  
  shown in Figure~\ref{fig:xib-signal}. 
  An unbinned likelihood fit finds~\cite{exp:xib-prl}  
  \( 17.5\pm4.3\stat\,\Xib \) candidates at a mass of  
  \( 5792.9\pm2.5\stat\pm1.7\syst\mevcc \) and with a significance of  
  \( 7.7 \) of Gaussian standard deviations.  The 
  results~\cite{exp:xib-prl} are in good agreement with theoretical 
  predictions and with the observation made by the 
  {D\O}~Collaboration~\cite{exp:xib-D0}. 
\section{ Correlated $\bbbar$ Production in \cdf2 Detector } 
\begin{figure}[hbt] 
  \hspace{-0.3in}\includegraphics[width=1.2\columnwidth,height=1.2\columnwidth]{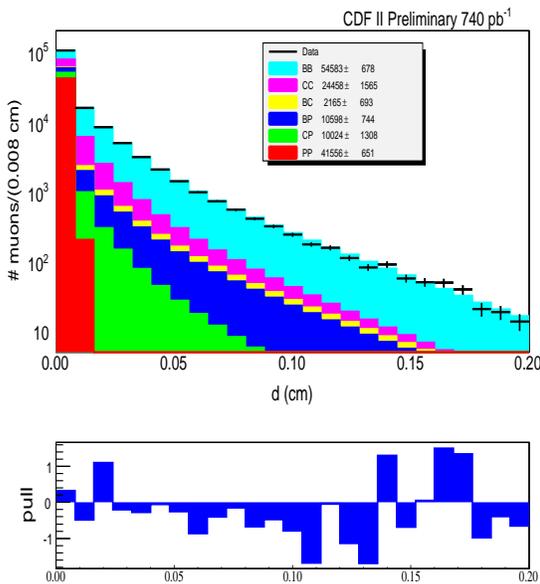} 
  \caption{The projection of the 2-dimensional fit of \( \IP(\mmu_{1}),\IP(\mmu_{2}) \) 
           with background templates summed up and data superimposed. The notations used are 
           ``B'' as \b-source, ``C'' as \c-source and ``P'' as the source of prompt muons.} 
  \label{fig:d0}  
  \end{figure}

  In this chapter we cover briefly a unique analysis on a paired 
  $\bbbar$ production measurement.  As leading order (LO) processes 
  dominate \bbbar production, $\sigma_{\bbbar}$, while next-to-leading 
  (NLO) processes are essential for inclusive $\sigma_{\b}$ studies, 
  the measurement of \( \sigma_{\bbbar} \) will help to disentangle LO 
  and NLO contributions and to resolve the controversy between the Run 
  I {D\O} and CDF measurements~\cite{exp:bb-cdf-d0}.  We select dimuon 
  events with invariant masses \( 5 <M(\mmu_{1}\mmu_{2})< 80\gevcc\), 
  outside of the domain populated by sequential decays of single \b- 
  quarks and \( Z^{0} \) modes, and extract  
  \( \sigma( \b\rightarrow\mun\,+X,\,\bbar\rightarrow\mup\,+X ) \), subtracting 
  contributions from \ccbar, prompt Drell-Yan pairs, \c- and \b- onium 
  prompt decays, $\pi$-, \kaon- decays, and misidentified dimuon 
  candidates. The signal and background contributions are determined 
  by fitting the experimental 2-dimensional impact parameter  
  \( \IP(\mmu_{1}),\IP(\mmu_{2}) \) distribution to corresponding 
  templates expected for various dimuon sources.  The method exploits 
  the fact that the shape of the $\IP(\mmu)$ distribution is largely 
  determined by the lifetime of its parent heavy hadron.  The analysis 
  is based on a data sample of total luminosity \( {\Lumi = 740\invpb} \)  
  collected with the CDF dimuon trigger~\cite{cdf:det} having no 
  biases with respect to $\IP(\mmu)$ distribution.  The projection of 
  the 2-dimensional fit onto $\IP(\mmu)$ comprising various background 
  contributions is shown in Figure~\ref{fig:d0}. 
  The extracted experimental cross-section is found to be  
  \( \sigma(\b\to\mun,\bbar\to\mup)= 1549\pm133 \pb \).  
 The exact NLO predictions are made using Herwig Monte-Carlo 
 program~\cite{th:herwig}, MNR code~\cite{th:MNR} running with 
 EVTGEN generator~\cite{th:evtgen}, parton structure functions from MRST~\cite{th:MRST} 
 fits and Peterson fragmentation function~\cite{th:peterson}. 
  The ratio of data to NLO theoretical Monte-Carlo calculation is found to be 
   \( R2(\b\to\mun,\bbar\to\mup) = 1.20\pm0.21 \). The errors include  
  statistical and systematic uncertainties added in quadratures. 
  From this measurement we derive  
  \( \sigma(\bbbar, \pt\geq 6\gevc, |y|\leq 1) = 1618\pm148 \nb \). 
  The systematic uncertainty due to choice of the fragmentation  
  model is \(\sim25\% \). 
\section{Summary}  
  CDF announces the first observation of four bottom baryon \Sgbstpm 
  resonance states.  CDF has also observed the strange bottom cascade 
  baryon \Xibm, and our measurements are in agreement with the {D\O} 
  observation and with theoretical predictions.  \cdf2 detector has 
  measured the correlated production cross-section of $\b\bbar$ pairs 
  with \b-quarks identified in their muonic semileptonic modes. The 
  measurement is consistent with theoretical expectations. Using NLO 
  Monte-Carlo cross-section calculations, the full $\bbbar$ production 
  cross-section in the kinematic domain  
  \( (\pt\geq 6\gevc, |y|\leq 1)\) has been derived. 
\section{Acknowledgments} 
  The author is grateful to his colleagues from the CDF $B$-Physics 
  Working Group for useful suggestions and comments made during 
  preparation of this talk. The author thanks S.~C.~Seidel for support 
  of this work. 
\begin{footnotesize} 
% 
% IF YOU DO NOT USE BIBTEX, USE THE FOLLOWING SAMPLE SCHEME FOR THE REFERENCES 
% 
% ---------------------------------------------------------------------------- 
  
% 
% ---------------------------------------------------------------------------- 
\end{footnotesize} 
 
% **************************************************************************** 
% END OF BIBLIOGRAPHY AREA 
% **************************************************************************** 
 
\end{document}